\documentstyle[aps,prl,twocolumn,psfig]{revtex}

\title{Modeling of Covalent Bonding in Solids 
by Inversion of Cohesive Energy Curves}
\author{Martin Z. Bazant and Efthimios Kaxiras}
\address{Department of Physics, Harvard University,
Cambridge, MA 02138}

\date{received 24 April 1996}

\begin{document}
\maketitle
\begin{abstract}
We provide a systematic test of empirical theories of covalent bonding
in solids using an exact procedure to invert {\it ab initio} cohesive
energy curves.  By considering multiple
structures of the same material, it is possible for the first time to
test competing angular functions, expose inconsistencies in the basic
assumption of a cluster expansion, and extract general features of
covalent bonding. We test our methods on
silicon, and provide the direct evidence that the Tersoff-type
bond order formalism correctly describes coordination dependence.
For bond-bending forces, we obtain
skewed angular functions that favor small angles, unlike existing
models.  As a proof-of-principle demonstration, we derive a Si
interatomic potential which exhibits comparable accuracy to existing
models.
\end{abstract}

PACS numbers: 61.50.Lt,34.20.Cf,33.15.Dj 
\vskip 12pt

	Large-scale atomistic simulations are becoming increasingly
important in the study of complex physical phenomena such as fracture,
plastic deformation, two- and three-dimensional melting, epitaxial
growth, shock wave propagation, friction, sintering, etc.  Ideally,
one would like to represent the atomic interactions in these
simulations with a quantum mechanical approach, treating explicitly
the electronic degrees of freedom.  This is a computationally
demanding proposition, tractable at present only for relatively small
system sizes of order $10^2$ atoms.  An alternative description is in
terms of effective interatomic potentials which allow fast evaluation
of energies and forces, making possible simulations involving more
than $10^8$ atoms.  The drawback in going from an explicit quantum
treatment of electrons to an effective interatomic potential is a
significant loss in accuracy that may undermine simulation results.
In most cases, the microscopic mechanisms of greatest interest (for
instance, bond formation or rupture) are precisely those which require
high degree of transferability, that is, ability of the potential to
describe accurately a wide range of local atomic environments.

	Over a decade of experience has shown that such
transferability is difficult to attain, especially in covalent solids,
for inherently quantum effects such as bond bending and breaking,
hybridization, charge transfer, and metalization.  In the prototypical
case of silicon, about thirty model potentials exist in the literature
\cite{balamane}, including popular and innovative ones by Stillinger
and Weber (SW) \cite{SW}, Tersoff \cite{tersoff}, and Chelikowsky {\it
et. al.} \cite{chelikowsky}.  Although the shortcomings of existing
model potentials have been carefully documented, it has proven very
difficult to improve them or to understand, even qualitatively, the
causes of their failures \cite{balamane,KP}.  Some theoretical
arguments have been advanced to motivate the form of an effective
interaction \cite{carlsson0,carlsson} and to derive potentials as
approximations of quantum models
\cite{abell,pettifor,fedders,mccarly}, but little specific theoretical
guidance exists to aid in the development of new potentials. The most
successful approach to date is to guess a functional form using
physical intuition and then adjust parameters to fit a database of
{\it ab initio} structural energies\cite{balamane}.  The reliance on
intuition and fitting leads to the two questions that motivate our
work: (1) Is there an {\it ab initio} justification for the
functional form of an interatomic potential?  and (2) Given a
particular form, is there a systematic way to obtain new potentials
directly from {\it ab initio} energy data?

	In this Letter, we present an exact procedure for inverting
{\it ab initio} energy data to obtain parameter-free many-body
potentials \cite{bazant}.  The inversion approach was pioneered by
Carlsson, Gelatt, and Ehrenreich (CGE) for the case of a {\em pair}
potential \cite{CGE}, and since then the same formula has been applied
with limited success by only a couple of authors \cite{carlsson,wang}.
We revisit the inversion approach with the following innovations: (i)
a recursive formulation that incorporates many-body interactions and
strains other than uniform volume expansion; (ii) the
requirement that {\it ab initio} energies be exactly reproduced for
relevant densities only (near the equilibrium solid
and liquid densities); and (iii) the use of an overdetermined set of
structures for the same material, which guarantees a wide range of
relative atomic arrangements.  These ideas form a general framework
for analyzing functional forms and deriving potentials, as illustrated
by application to Si.  In this manner, we provide satisfactory answers
to both questions posed in the previous paragraph.  We analyze the two
defining features of covalent bonding in two steps, {\em 1. Pair
bonding} and {\em 2. Angular forces}.

{\em 1. Pair bonding:} 
We begin with the simplest case of a pair potential, in which
the cohesive energy $E[\phi]$ of an arbitrary structure is given by,
\begin{equation}
E(r) = \sum_{i \neq j} \phi( R_{ij} ) =  \sum_{p=1}
^{\infty} n_p \phi 
( s_p r ) , 
\label{eq:ephi}
\end{equation}
with atomic separations grouped into shells $S_p$ of radius $s_p r$
containing $n_p$ atoms each. Dilation of the lattice is achieved by
varying the parameter $r$ with the structural quantities $\{s_p\}$ and
$\{n_p\}$ fixed.  Shells are numbered so that $s_1 <
s_2 < s_3 < \ldots$, and distances scaled so that 
$s_1 = 1$.  A simple rearrangement of the terms in Eq. (\ref{eq:ephi})
yields the desired inversion formula for $\phi[E]$,
\begin{equation}
\phi(r) = \frac{1}{n_1} \left( E(r) -
        \sum_{p=2}^\infty n_p \phi( s_p r) \right)  .
\label{eq:phirec}
\end{equation}
Although the unknown potential appears on both sides of this equation,
recursive substitution generates the explicit formula,
\begin{equation}
\phi(r) =  \frac{1}{n_1} E(r) - \sum_{p=2}^\infty
           \frac{n_p}{n_1^2} E(s_p r) 
        + \sum_{p,q = 2}^\infty  \frac{n_p n_q}{n_1^3}
        E(s_p s_q r) - \ldots  ,
\label{eq:phiformula}
\end{equation}
which was originally derived by CGE invoking the linearity of the
functional $E[\phi]$ \cite{CGE}.  Our recursive formulation
generalizes to nonlinear functionals and suggests a simple
computational procedure: If the tail of $\phi(r)$ is assumed known for
$r>a$, then the potential is uniquely determined by solving the
recursion in order of decreasing $r$ starting at $r=a$ (because $s_p >
1$ for $p \ge 2$).  An important case is that of finite range,
i.e. $\phi(r)=0$ for $r \ge a$, as is 
typically assumed for interatomic potentials.

\begin{figure}
\begin{center}
\mbox{
\psfig{file=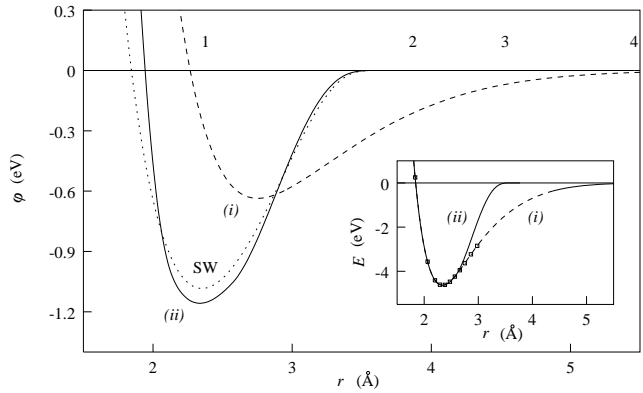,height=2in}
}
\end{center}
\caption{The inverted pair potential for silicon (i) before and (ii)
after a cutoff is imposed, compared with $\phi_{SW}(r)$ (dashed line).
Numbers inside the figure indicate shell radii in the diamond lattice. The
inset shows the diamond LDA data and the interpolant (i) before and
(ii) after imposing a cutoff.}
\end{figure}

	To illustrate the inversion procedure, we apply it to an {\it
ab initio} database consisting of cohesive energy curves for Si
from density functional calculations in the local density
approximation (LDA) \cite{LDA}. 
In order to keep the procedure simple while still
capturing the important local bonding characteristics, the database includes
the following crystals:
(i) the low-energy and low-coordination structures, diamond (Si-I),
$\beta$-tin (Si-II) \cite{cohen}, BC-8 (Si-III) \cite{yin}, and BCT-5
\cite{boyer}; (ii) SC and FCC crystals for metallic behavior; and
(iii) the graphitic structure for non-tetrahedral hybridization
\cite{granote}. These structures have coordinations 4,6,4,5,6,12, and
3, respectively. We consider only atomic volumes smaller than
$(3.54\AA)^3$ to avoid the difficulty of LDA to represent accurately
the energies of isolated atoms \cite{moll}. Smooth interpolation of
the LDA data and extrapolation to infinite volume with an exponential
tail are used.  The LDA data points for the diamond lattice with the
interpolant are shown in the inset of Fig. 1.

	The inverted pair potential for the diamond curve, shown in
Fig. 1, is clearly unphysical: Its long range and strong repulsion at
the first neighbor distance contradict our intuitive understanding of
covalent bonding.  Similar results have been obtained in previous work
applying the CGE formula to metals \cite{carlsson,CGE} and semiconductors
\cite{wang}.  Our recursive approach reveals that these problems are
inherent to the inversion process, which, in spite of being exact,
stretches the assumption of a volume-independent potential to an
unphysical extreme.  Because inversion amounts to solving in order of
decreasing distance from infinite separation, the tail of the
potential comes from unscreened interactions between atoms in a low
density gaseous phase \cite{carlnote}.
The same tail is then used to
describe long range interactions in a bulk crystal, which are
presumably screened by the presence of closer atoms. While the nature
of screening and its description by an effective potential are
subjects of active research, it is obvious that distant atoms in a
bulk crystal cannot interact in the same way as atoms with the
same separation in a gas.

\begin{figure}
\begin{center}
\mbox{
\psfig{file=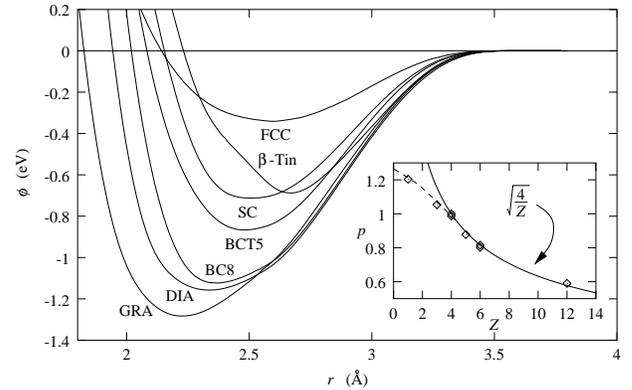,height=2in}
}
\end{center}
\caption{Inverted pair potentials (with cutoff) for seven silicon
bulk phases. The inset shows the implied bond order $p$ extracted from
these curves (points) compared to $\sqrt{4/Z}$ (line). $p(1)$
reflects the Si$_2$ bond length and energy \protect\cite{balamane}.
}
\end{figure}

	To rectify the inversion procedure, we forgo the requirement
that the potential exactly reproduce the {\em entire} cohesive energy
curve.  Instead, we focus on condensed volumes typical of solid and
liquid environments, whose exact energies can be preserved with any
choice of tail for the potential. For Si, we find that
exponential decay to a cutoff near the second-neighbor distance in the
diamond lattice, $3.84 \AA$, generates potentials in good agreement
with bonding theory. For example, as shown in Fig. 1,
we force the energy curve to be zero for $r \geq a_{SW} = 3.77118 \AA$
without disturbing energies within 10\% of the equilibrium bond
length, where covalent bonds are well-defined, to produce an inverted
pair potential with a deep minimum at the first neighbor distance.

	Applying the same procedure to the other curves in our
database, we obtain the potentials of Fig. 2. The large discrepancy
between them is direct evidence for a well-known fact: the energetics
of silicon {\em cannot be described by a pair potential alone}
\cite{carlsson}. Our results suggest, however, that 
an environment-dependent pair potential
can describe the ideal bulk
phases reasonably well. There is a clear coordination dependence to
the curves: bond lengths (positions of the minima) increase, and bond
strengths (depths of the minima) decrease with increasing
coordination.  This behavior can be described by the
bond-order formalism, which is justified on grounds of theoretical
arguments \cite{carlsson0,carlsson,abell,pettifor} as well as
experience with empirical potentials \cite{tersoff,dodson,bolding}. In
its simplest form, a bond order potential is given by,
\begin{equation}
\phi(r,Z) = \phi_R(r) + p(Z) \phi_A(r) ,
\label{eq:pz}
\end{equation}
where $\phi_R$ and $\phi_A$ are monotonic repulsive and attractive
terms, respectively, and $p(Z)$ gives the bond strength as a function
of the coordination $Z$. The leading order approximation of the bond
order is $p(Z) \propto Z^{-1/2}$. Since this comes from describing the
local density of states by the bandwidth only, we expect the
approximation to work well for the metallic phases with $Z > 4$
(BCT-5, $\beta$-tin, SC, and FCC). For the covalent phases with $Z
\le 4$ (diamond, BC-8, and graphite), band shape effects become
important and we expect significant departure from the $Z^{-1/2}$
behavior.  

	If the repulsive interaction $\phi_R$ were known, the
bond-order term could be extracted directly from the {\it ab initio}
data, using $p(Z) = V_A(r_0) / V_A^{dia}(r_0)$, where $V_A = \phi -
\phi_R$, $r_0$ is the minimum of the inverted potential $\phi$, and we
set $p=1$ for the diamond lattice ($Z = 4$).  The repulsive term,
intended to represent an effective force between electrons due to
Pauli exclusion, is the weakest link in bond-order models, since its
form must be assumed and then fit to empirical data without
theoretical guidance.  Although the general trend is insensitive to
the choice of $\phi_R$, we find that using $\phi_R = 2 \phi_R^{SW}$,
where $\phi_R^{SW}$ is the repulsive part of the SW potential,
produces a $p(Z)$ which lies remarkably close to its expected behavior
(see inset of Fig. 2).

{\em 2. Angular forces:}
	While the energetics of bulk phases can be fairly well
described by a bond order pair functional, it is well-known that
many-body interactions with explicit angular dependence are required
for silicon, for example, to stabilize the diamond lattice against
shear strain \cite{tersoff,carlsson}. As the simplest case of a
many-body potential, we consider one with volume-independent
pair terms and separable three-body terms like the potentials of SW and
Kaxiras and Pandey \cite{KP}.  The many-body energy, $F(r) = E(r) -
V_2(r)$, formed by subtracting the pair terms $V_2(r)$ from
the total energy $E(r)$, is expressed as a sum over pairs of bonds,
\begin{equation}
F(r) = \sum_i \sum_{j \neq i} \sum_{k \neq i, k > j}
       g( R_{ij} ) g( R_{jk} ) h(\theta_{ijk}) ,
\label{eq:gexpansion}
\end{equation}
where $\cos \theta_{ijk} = \hat{R}_{ij} \cdot \hat{R}_{jk}$. Following
theory \cite{carlsson0,carlsson} and practice \cite{balamane}, we
assume $F \ge 0$, which implies that the pair potential come from the
diamond lattice inversion described above.  A particular form for the
angular term $h(\theta)$ must be assumed in order to invert $F[g,h]$
for the radial function $g[F,h]$.  The procedure is the same as in the
pair potential case: solve Eq. (\ref{eq:gexpansion}) for $g(r)$ to
obtain a recursion. Grouping bonds into shells as above and taking the
positive root of the resulting quadratic equation yields the desired
expression,
\begin{equation}
g(r) = \frac{ -\beta(r) + \sqrt{ \beta(r)^2 + 4 \alpha_{11}
    ( F(r) - \gamma(r) ) }}{ 2 \alpha_{11} } ,
\label{eq:grecursion}
\end{equation}
where $\alpha_{pq} = \sum_{r_{ij} \in S_p} \sum_{r_{ik} \in S_q}
h(\theta_{ijk})$, \ $\beta(r) = \sum_{p=2}^\infty \alpha_{1p} g(s_p r)$,
and $\gamma(r) = \sum_{p=2}^\infty \sum_{q=p}^\infty \alpha_{pq} g(s_p
r) g(s_q r)$.  In the $\alpha_{pp}$ sums, only $k>j$ contributes to
avoid double counting.  An explicit formula like the CGE pair
potential can be obtained by recursive substitution with
Eq. (\ref{eq:grecursion}) and involves a tree of nested square roots.
It is simpler to follow the same computational procedure as before,
solving the recursion in order of decreasing distance starting at the
cutoff.  In principle, the same general procedure can be applied to
determine radial functions of other forms or for higher order terms in
a cluster expansion of the effective potential.  For example, for a
nonseparable three-body term involving three bond lengths, like the
potential of Pearson et al. \cite{PTHT}, the recursion comes from
solving a cubic equation, and for a four-body term, a quartic
equation.

\begin{figure}
\begin{center}
\mbox{
\psfig{file=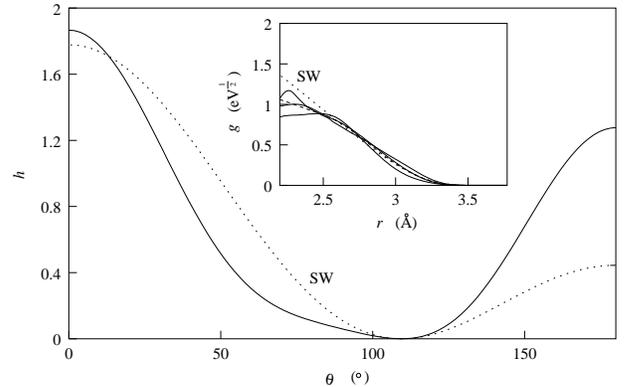,height=2in}
}
\end{center}
\caption{An inverted angular function for silicon from the BC-8,
BCT-5, and $\beta$-tin energy curves compared with $h_{SW}(\theta)$.
The inset shows the collapse of the inverted $g(r)$
with the average curve (dashed line) and $g_{SW}(r)$ (dotted line).}
\end{figure}

	As in the pair potential case, it is useful to invert more
than one cohesive energy curve of the same material for $g[F,h](r)$.
If the assumed angular dependence $h(\theta)$ (and two-body terms)
were truly transferable, then the same radial function $g(r)$ would
result from every inversion. Conversely, the greater the variance
between the inverted $g(r)$, the less transferable is the assumed
$h(\theta)$.  This principle gives us a {\it quantitative} means of
assessing the quality of angular functions directly from the {\it ab
initio} data.  For example, although the SW angular function
$h(\theta) = (\cos(\theta) + \frac{1}{3})^2$, produces mediocre
collapse of the inverted $g(r)$, it is fortuitously far better than taking
$h(\theta) = (\cos(\theta) + \frac{1}{3})^4$, due to the latter's
flatness near the tetrahedral minimum. 

	Using the same principle, we can extract an optimal angular
dependence from the {\it ab initio} data by assuming a series
expansion, $h(\theta) = \sum_{i=0}^2 c_i ( \cos(\theta) +
\frac{1}{3})^{2+i}$. For the curve shown in Fig. 3 (defined by $c_0=1,
c_1=-1.86, c_2=1.42$), the collapse of radial functions from the low
energy phases is rather good, as seen in the inset of Fig. 3. A novel
feature of the inverted angular function is its skew about the minimum
to favor
smaller angles. This is consistent with the conclusion that
existing potentials tend to overpenalize angles smaller than $\pi/2$
\cite{balamane}, which presumably leads to poor descriptions of
surfaces, clusters, and certain defects. The skewed angular function
also raises the energy of overcoordinated metallic structures relative
to covalent ones by penalizing large angles. While it is typical to
characterize metallic structures by the presence of small angles
\cite{chelikowsky}, we note that metallic structures tend to have
angles near $\pi$ also. Covalent bonds are actually characterized by
angles in the intermediate range, $\pi/2$ to $2 \pi/3$.
 
 	Although we are not attempting here to provide an improved
potential for Si, we have performed some tests of the potential
obtained by the inversion just described \cite{potnote} 
(in this proof-of-principle
demonstration, we omit coordination dependence for practical reasons). 
We find that it performs as well as the popular SW and Tersoff
potentials without fitting to any defect structures, for energies of
other silicon bulk phases, defects such as interstitials and
vacancies, generalized stacking faults, the concerted exchange
diffusion mechanism \cite{KP}, and (100) and (111) surface
reconstructions.  Considering the database employed in the inversion,
we conclude that many important features of chemical bonding are
contained in cohesive energy curves for ideal bulk phases.

	In conclusion, we have presented a general procedure for
inverting cohesive energy curves to obtain many-body effective
interatomic potentials. By inverting {\it ab initio} cohesive energy
curves for silicon, we have demonstrated how general features of
bonding are revealed.  Elsewhere we will describe extensions of these
ideas, for example, to the inversion of energy curves for shear
strains to obtain the angular function $h[F,g](\theta)$ directly. The
inversion procedure provides a systematic method for deriving
interatomic potentials and a unique tool for understanding their
general limitations through the direct use of {\it ab initio} data.
It is hoped that this tool will lead to potentials with improved
transferability, a goal that has proven elusive when pursued by
intuitive arguments and fitting of databases.

	MZB acknowledges a Computational Science Graduate Fellowship
from the Office of Scientific Computing of the Department of Energy.
This work was supported in part by ONR grant \#N00014-93-I-0190.

\end{document}